\documentstyle[aps,twocolumn,prl,epsf]{revtex}

\newcommand{\be}{\begin{equation}}
\newcommand{\ee}{\end{equation}}
\newcommand{\bea}{\begin{eqnarray}}
\newcommand{\eea}{\end{eqnarray}}

\begin{document}
\draft

\twocolumn[\hsize\textwidth\columnwidth\hsize\csname
@twocolumnfalse\endcsname
\title{Phase Diagram of the Two-Channel Kondo Lattice}
\author{
Mark Jarrell$^{(a)}$, Hanbin Pang$^{(a)}$, and D.L.\ Cox$^{(b)}$
}
\address{
$^{(a)}$ Department of Physics,
University of Cincinnati, Cincinnati, OH 45221\\
$^{(b)}$Department of Physics,
The Ohio State University, Columbus, OH, 43202 \\
}
\date{\today}
\maketitle

\widetext
\begin{abstract}
\noindent
The phase diagram of the two-channel Kondo lattice model is examined
with
a Quantum Monte Carlo simulation in the limit of infinite dimensions.
Commensurate (and incommensurate) antiferromagnetic and superconducting
states are found.
The antiferromagnetic
transition is very weak and continuous; whereas the superconducting
transition
is discontinuous to an odd-frequency channel-singlet and spin-singlet
pairing
state.
 \end{abstract}

 ]

 \narrowtext
\paragraph*{Introduction.}

A number of Heavy Fermion materials display highly unusual
superconductivity (HFSCs)\cite{ott}. It seems unlikely
that conventional superconductivity
coexists with the strong local Coulomb correlations necessary to
enhance
the electronic mass.  Indeed, the specific heat jump at the transition
scales with the normal state specific heat making it clear pairing is
between
the heavy electrons (with order 100-1000-fold mass enhancement).  The
strong coulomb correlations  present no problem for unconventional
superconducting
order parameters with either spatial\cite{sauls} or temporal
nodes (so
called ``odd-frequency
pairing)\cite{oddfreq,comirtsv,emkiv,cft_long,zachemkiv}.
Such an interpretation is supported by two sets of data: (i) power laws
observed in
physical properties below the superconducting
transition\cite{ott,sauls}, which
contrast with the activated behavior of the conventional (nodeless)
$s$-wave order
 of, e.g.,  aluminum; (ii) the complex superconducting
phase diagrams of UPt$_3$ and U$_{1-x}$Th$_x$Be$_{13}$ (and possibly
UBe$_{13}$ itself,
for which the penetration depth displays evidence of a secondary
transition)\cite{ott,sauls,cox-pair}.
The superconductivity can coexist with antiferromagnetism, which is
usually commensurate, as in UPt$_3$, URu$_2$Si$_2$, UPd$_2$Al$_3$,
and UNi$_2$Al$_3$\cite{ott}, or compete with it as in
CeCu$_2$Si$_2$\cite{steggoa}.
Finally, at least in UBe$_{13}$\cite{ottold} and
CeCu$_2$Si$_2$\cite{steggoa}, the
superconductivity arises in a normal state which is clearly {\it not}
described as
a Fermi liquid.
In each of these materials above $T_c$
the linear specific heat
rises with decreasing temperature, the resistivity is approximately
linear
in $T$, and the residual resistivity at $T_c$ is high (typically
80-100 $\mu-\Omega$-cm in the best samples of UBe$_{13}$).

In this paper, we provide the first calculations of the phase diagram
for the
two-channel Kondo lattice in infinite spatial dimensions.  We find
second order
antiferromagnetic and first order odd frequency superconducting phase
transitions.
Coexistence of commensurate (and incommensurate)
antiferromagnetism with superconductivity is in principle possible.
We present several possible routes to account for the multiple
superconducting
phases observed in real materials.  Taken together with the non-Fermi
liquid
(NFL) paramagnetic phase discussed in previous
publications\cite{twochannlatt},
and earlier suggestions that the two-channel lattice may describe
UBe$_{13}$ and
other heavy fermion systems\cite{cox-pair}, our work establishes the
two-channel Kondo lattice model as possessing the
key ingredients needed to explain heavy fermion superconductivity.

\paragraph*{Motivation} A possible model for some heavy
fermion compounds is the two-channel Kondo lattice which consists of
two identical species of non-interacting electrons
antiferromagnetically
coupled to an adjacent lattice of spin 1/2 Kondo
moments\cite{cox-pair}.
This model displays Non-Fermi liquid behavior because of the
overcompensation
of the  Kondo spins by the conduction electrons, first pointed out in
the single
site two-channel model by Nozi\`{e}res and Blandin\cite{nozbland} where
extended
screening clouds form which retain a spin 1/2  character. As a result,
the
paramagnetic state is degenerate and the excitation spectrum non-Fermi
liquid like.
This overscreening results in an interchannel pairing
mechanism\cite{cox-pair}
in the two-channel impurity problem which tends to favor spin-singlet
channel-singlet
{\it odd frequency}
superconducting fluctuations\cite{emkiv,cft_long,zachemkiv}.
For the lattice, we find that
this overscreening generates
novel antiferromagnetic superexchange between the Kondo
spins which, taken
 together with the RKKY exchange, favors antiferromagnetism
in the lattice model close to half
filling of the bands.
The nature of the resulting superconducting and magnetic
transitions, as well
as the competition between them, is the subject of this manuscript.
 .

\paragraph*{Model} The Hamiltonian for the two-channel Kondo
lattice is
$$H=J\sum_{i,\alpha}{\bf{S}}_i\cdot {\bf{s}}_{i,\alpha}
    -{t^*\over 2\sqrt{d}}\sum_{<ij>,\alpha,\sigma}
    \left(c_{i,\alpha,\sigma}^{\dag} c_{j,\alpha,\sigma}
    +\mbox{h.c.}\right) $$
\begin{equation}
    -\mu\sum_{i,\alpha,\sigma}c_{i,\alpha,\sigma}^{\dag}
    c_{i,\alpha,\sigma}\,,
    \label{Ham}
\end{equation}
where $c_{i,\alpha,\sigma}^{\dag}$ ($c_{i,\alpha,\sigma}$)
creates (destroys) an electron on site $i$ in channel $\alpha=1,2$
of spin $\sigma$, ${\bf{S}}_i$ is the Kondo spin on site $i$, and
${\bf{s}}_{i,\alpha}$ are the conduction electron spin operators for
site
$i$ and channel $\alpha$.  The sites $i$ form an infinite-dimensional
hypercubic lattice.  Hopping is limited to nearest neighbors with
hopping integral $t\equiv t^*/2\sqrt{d}$; the scaled hopping integral
$t^*$
determines the energy unit and is set equal to one $(t^*=1)$.  Thus, on
each
site the Kondo spin mediates an interaction between the two different
channels.  This problem is non-trivial, and for the region of interest
in
which $J>0$ and $T\ll J$, $t^*$ it is describable only with
non-perturbative
approaches.   Clearly some simplifying method which allows for a
solution of the lattice problem in a non-trivial limit is necessary.

\paragraph*{Formalism and Simulation.}
Such a method was proposed by Metzner and Vollhardt \cite{mevoll} who
observed
that the renormalizations due to local two-particle interactions become
purely
local as the coordination number of the lattice increases.
 A consequence is that the solution of most standard lattice
models may be mapped onto the solution of a local correlated system
coupled
to an effective bath that is self-consistently
determined.  We
refer the reader to recent reviews for further
details and references on the method\cite{infdrev}.

 In order to solve the remaining impurity problem, we use the Kondo
impurity algorithm of Fye and Hirsch\cite{fye}, modified to simulate
the
two-channel problem\cite{luk}.
 We simulated the model for a variety of fillings and values of the
magnetic exchange $J$ ($ 0< N\leq1$ for $J=0.75, 0.625, 0.5, 0.4$).
Error bars on measured quantities
are less than 6\% for the results presented here.  A sign problem was
also encountered in the QMC process.  It becomes severe
for large $J$ and also for low fillings since lower temperatures are
required
to access the physically interesting regime.  This precludes a
systematic
study of the model with $J>0.8$ and filling $N<0.5$.

	Both one and two-particle properties are natural products
of the QMC simulation.  To calculate the spin susceptibility, it
was necessary to measure the local three-by-three matrix of the
susceptibility including both the Kondo spin and conduction band
spin fluctuations.  This may then be inverted to calculate the
associated
irreducible vertex function and the corresponding lattice
susceptibility
in the usual way\cite{jarrell_pam}.

The situation for
the superconductivity is a bit more complicated.  Since there
are no charge fluctuations involving the Kondo spin degree of
freedom, only the two conduction channels contribute to the pair-field
susceptibility.  We can then look for pairing instabilities in singlet
and triplet
channels for both spin and channel spin.
Motivated by the result for the impurity model\cite{emkiv,cft_long}
 we have restricted our attention to interchannel and interspin
 particle-particle
diagrams as shown in Fig.~1.
It is possible to make two independent combinations of these diagrams
$\chi_{\pm}(i\omega_n,i\omega_m)=\chi_{11}(i\omega_n,i\omega_m) \pm
\chi_{12}(i\omega_n,i\omega_m)$,
from which we construct a quartet of spin and channel, singlet and
triplet pair-field susceptibilities given by
\begin{eqnarray}
P_{SsCs} & =
& T\sum_{nm} f_-(i\omega_n)\chi_-(i\omega_n,i\omega_m)f_-(i\omega_m)\\
P_{StCt} & =
& T\sum_{nm} f_+(i\omega_n)\chi_+(i\omega_n,i\omega_m)f_+(i\omega_m)\\
P_{StCs} & = & T\sum_{nm} \chi_-(i\omega_n,i\omega_m)\\
P_{SsCt} & = & T\sum_{nm} \chi_+(i\omega_n,i\omega_m)\,.
\end{eqnarray}
Here $f_\pm(i\omega_n)$ are odd functions of Matsubara frequency used
to project out the odd-frequency pairing, and, for example,
$P_{SsCs}(T)$
is the spin-singlet channel-singlet pair-field susceptibility.

	To determine the form of $f_{\pm}(i\omega_n)$, we employ the
	pairing matrix
formalism of Owen and Scalapino\cite{owen_scal}.  Here we represent
each of
$\chi_{\pm}$ in a two-particle Dyson equation and extract the
irreducible
vertex functions $\Gamma_{\pm}$.  The pairing matrices are then
\begin{equation}
M_{\pm}(i\omega_n,i\omega_m)=
\sqrt{\chi_{\pm}^0(i\omega_n)}
\Gamma_{\pm}(i\omega_n,i\omega_m)
\sqrt{\chi_{\pm}^0(i\omega_m)}\,.
\end{equation}
where $\chi_{\pm}^0(i\omega_n)$ are the particle-particle diagrams
in Fig.~1 without vertex corrections.
$f_\pm(i\omega_n)$ is the eigenvector corresponding to the dominant
eigenvalue of $M_\pm$ (that with the largest absolute value),

\paragraph*{Results.}
In this model antiferromagnetism is driven by both RKKY interactions
 and a novel type of
superexchange.  The latter arises from hopping between adjacent
spin 1/2 screening clouds, whose overall spin is determined by the
conduction
electrons; the Pauli principle forbids hopping unless neighboring spins
in
the same channel are antiparallel. As a result, for large $J$, the
superexchange goes as $\sim (t^*)^2/J$.
 For conduction band fillings close to $N=1$ both
the RKKY (evaluated at nearest neighbor sites)
and the superexchange favor antiferromagnetism (the RKKY
exchange remains antiferromagnetic until $N\alt 0.5$), and an
antiferromagnetic
transition results, as shown in Fig.~2.  Due to screening of the local
moments by the conduction spin, the transition is very weak, as
measured by the
full susceptibility.
Specifically, $\chi_{AF}$ is not significantly enhanced over the bulk
susceptibility
$\chi_F$ until $T\agt T_N$.
However, the f-electron contribution to the susceptibility shows
a protracted scaling region.  Here
screening effects non-linear feedback that reduces the susceptibility
exponent $\gamma$ from the
mean field value $\gamma=1$.  $\gamma$ increases with doping ($N<1$),
and the transition becomes incommensurate as $T_N\to0$.

To explore superconductivity, it is necessary to find the frequency
form
factors $f_\pm$, which, as discussed above, are the eigenvectors
corresponding
to the dominant eigenvalues of the pairing matrices $M_\pm$.  As the
temperature
is lowered, the dominant eigenvalue first becomes large (divergent) and
negative, and then abruptly switches to a large and positive value at
the
transition.
 This happens first in $M_-$, and the corresponding eigenvector
of $M_-$ is  plotted in the inset to Fig.~3.  It can be fit quite
accurately to the form $T/2\omega_n$ as shown by the solid line, which
corresponds
to the form factor of Ref. \cite{comirtsv}.  Thus,
we use $f_-(i\omega_n)=T/2\omega_n$ to project out the odd-frequency
pair-field susceptibilities shown in Figs.~3 and 4.  (Other form
factors $f(i\omega_n)=\tanh(T_0/\omega_n)$ and $sign(\omega_n)$
were tried and produce qualitatively similar results.)

The first transition is found in the spin-singlet channel-singlet
pairing
combination, as shown in Fig.~3. (Note that this pair state is of
necessity
even in parity, so that the odd-frequency condition is required to
satisfy the
Pauli principle, in contrast with the spin-parity selection rules
in the single channel case\cite{oddfreq,comirtsv}.)
 To interpret this result, remember that
the inverse pair field susceptibility is proportional to the curvature
of
the free energy $f(\Delta_{SsCs})$ as a function of the pairing order
parameter $1/P_{SsCs} \propto d^2f(\Delta_{SsCs})/d\Delta_{SsCs}^2$.
Thus,
if $P_{SsCs}<0$ a thermodynamic instability of the system is present.
The associated transition cannot be continuous, since this requires the
free energy
in order parameter space to become flat (ie. $P_{SsCs}$ diverges) so
that the order parameter
may change continuously.  Thus, we identify this as a discontinuous
transition.  Furthermore, if $P_{SsCs}(T^*)=0$, then $T^*$ is a lower
bound
to the transition temperature since when $P_{SsCs}(T-T^*=0^-)=0^-$
the curvature of $f(\Delta_{SsCs})$ is divergent and negative. I.e.
the free energy displays a downward cusp which would compel the order
parameter and the free energy to change discontinuously at $T^*$, which
involves an infinite energy at the transition. Hence,
the actual transition occurs at a temperature $T_c>T^*$.

 Several remarks are in order about this unusual superconductivity: (a)
Fig.~3 suggests that the transition is degenerate at the zone center
and the zone corner, and in fact $P_{SsCs}$ vanishes simultaneously
over the whole zone, and hence this is a {\it locally} driven
transition!
This degeneracy may be lifted when the order parameter becomes finite,
otherwise, it will be lifted in finite dimensionality as the
superfluid stiffness $\Upsilon_s$ (analogous to spin stiffness in a
magnet)
must vanish in infinite $d$. For finite $d$ with $\Upsilon_s>0(<0)$, a
local free energy minimum
will be found at $\vec q=0(\vec q\ne 0)$. (b) The vanishing of the
local pair susceptibility here at $T^*$ contrasts with the impurity
model in which it diverges logarithmically as $T\to
0$\cite{emkiv,cft_long}. (c) The excitation spectrum of such a
transition may well be
highly exotic\cite{comirtsv}.

As shown in Fig.~4, the ground state of the system may be
superconducting
or magnetic which may coexist or compete. Our work gives only estimates
of
the transition temperatures; detailed exploration within the ordered
phases
will answer this question definitively.  In general superconductivity
will
occur first (at the highest transition temperature) away from half
filling
and antiferromagnetism will occur first near half filling.  However,
for
values of $J>0.75$, we find that superconductivity occurs first even at
half filling (the minus sign problem precluded a systematic study for
these
values of $J$ so they are not presented in Fig.~4).  We also find that,
$T^*\propto T_0$, with the constant of proportionality $\approx 0.5$
weakly
dependent on $J$; whereas $T_N$ appears to depend upon both $J$ and
$T_0$.

\paragraph*{Speculation and Interpretation}  We offer two compatible
interpretations of the result $T^*\approx 0.5 T_0$.  First, consistent
with the fact that the transition is locally driven, the transition
temperature is proportional to the Kondo scale $T_0$, the temperature
characteristic of the local dynamics.  Second, $T\approx 0.5T_0$ is
also
the temperature at which the slope of the real part of the self energy
becomes one\cite{twochannlatt}, so that the quasiparticle
renormalization
factor diverges.  This result suggests that when the system cannot form
a
Fermi liquid due to a large residual scattering rate (which also yields
a
large residual entropy), it forms a superconducting state to quench the
entropy.  The antiferromagnetic transition temperature, on the other
hand, depends upon both $J$ (through the intersite exchange) and upon
$T_0$ (through moment screening and superexchange).
 As noted above,
we cannot determine whether these states coexist
without a detailed exploration within the ordered phases. At half
filling, where
the antiferromagnetism produces an insulating phase, clearly the
superconductivity
will be suppressed for $J\le 0.75$.

Our results offer a number of routes to be explored for explaining
the complex superconducting phase diagrams of UPt$_3$ and
U$_{1-x}$Th$_x$Be$_{13}$:
 1) {\bf Competition between phases with a multi-point irreducible
 star}.
As mentioned above, for $\Upsilon_s<0$ in finite $d$, $\vec q\ne 0$
pairs are favored.  In
this simple case of a bipartite lattice, the zone corner has a
one-point
irreducible star.  For lattices with frustration, such as
hexagonal UPt$_3$ and face-centered cubic UBe$_{13}$,
it is possible to produce
multi-point irreducible stars for staggered order parameters
which can then have multiple phases\cite{heidetal}.
 2) {\bf Competition Between Different $\vec q$ values}.  As noted
 above,
$T^*$, the lower bound for the first order superconducting transition
temperature
is independent of $\vec q$ in our calculations.  Thus multiple phases
may thus
correspond to superconducting transitions with different $\vec q$
values.
3){\bf Possible Instability of Spin Triplet/Channel Triplet Pairing}.
At yet
lower temperatures than those identified in Fig.~4, we observe a sign
change in
the pair field susceptibility associated with spin triplet-channel
triplet
odd-frequency pairing. Hence it is possible that the competition
between this
triplet-triplet and the singlet-singlet odd frequency pairing may
explain the complex phase diagrams.

In closing, we note
that this superconducting transition can agree with experiment
only if it is weakly first order; this is plausible given the rapid
change in
free energy curvature at the order parameter origin.  Detailed
investigations
in the ordered phase will resolve this issue and whether any of the
above
scenarios can describe the complex phase diagrams of UPt$_3$ and
U$_{1-x}$Th$_x$Be$_{13}$.

 We would like to acknowledge useful discussions with F.\ Anders, A.V.\
Balatsky, W.\ Chung, A.\ Georges, B.\ Goodman, D.\ Hess, H.R.
Krishna-murthy,
M.\ Ma, A.J.\ Millis, and W.\ Putikka.  Jarrell and Pang would like to
acknowledge
the support of NSF grants DMR-9406678 and DMR-9357199.  Cox
acknowledges
the support of the NSF under Grant No. DMR-9420920. Computer support
was
provided by the Ohio Supercomputer Center.

\begin{figure}[htb]
\epsfxsize=3.8in
\epsffile{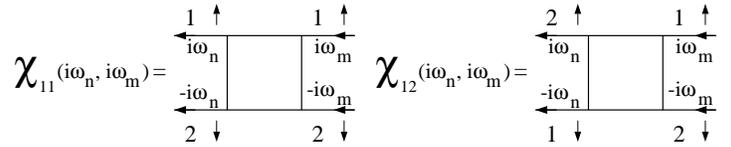}
\caption{Particle-particle interchannel opposite spin diagrams which
contribute to the pair-field susceptibility.  Here $1$ and $2$ label
the channel and $\uparrow$ and $\downarrow$ the spin.}
\end{figure}

\begin{figure}[htb]
\epsfxsize=3.8in
\epsffile{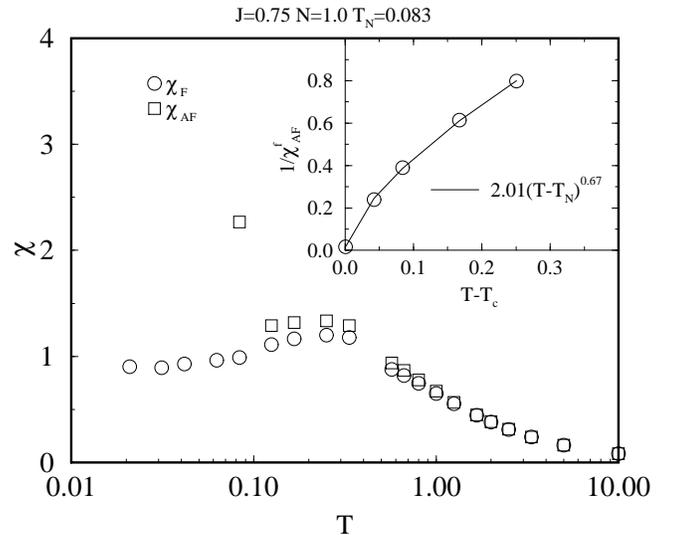}
\caption{{Full antiferromagnetic $\chi_{AF}$ and ferromagnetic
$\chi_{F}$
susceptibilities of the two-channel Kondo lattice with $J=0.75$ and
$N=1.0$.
$\chi_{AF}$ shows very little enhancement until $T\approx T_N$.  As
shown in
the inset, the Kondo spin contribution to the susceptibility
$\chi_{AF}^f(T)$
displays a protracted scaling region.  The solid
line is a  fit of $1/\chi_{AF}^f(T)$ to the form $a(T-T_N)^\gamma$. The
large difference between $\chi_{AF}$ and $\chi_{AF}^f(T)$ as well
as the reduced exponent $\gamma<1$ indicate the importance of screening
of the local spin in this transition.}}
\end{figure}

 \begin{figure}[htb]
\epsfxsize=3.8in
\epsffile{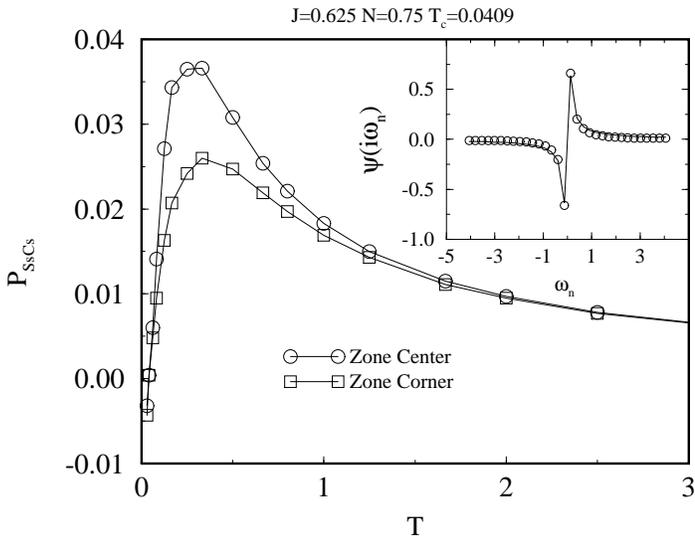}
\caption[]{{Odd-frequency (spin-singlet, channel-singlet) pair-field
susceptibility at both the zone center and corner.
At low temperatures $T=T^*\leq T_c$ $P_{SsCs}(T)$ becomes negative,
indicating
a thermodynamic instability to the formation of a pairing state.
The lack of a divergence of $P_{SsCs}(T)$ rules out a continuous
transition,
hence the system must undergo a discontinuous transition.
In the inset, the dominant eigenvector of the pairing matrix $M_-$ is
plotted versus Matsubara frequency.  It can be fit to the form
$T/2\omega_n$ (solid line), which corresponds to the frequency
variation
of the gap function found in Ref.4 }}
\end{figure}

 \begin{figure}[htb]
\epsfxsize=3.8in
\epsffile{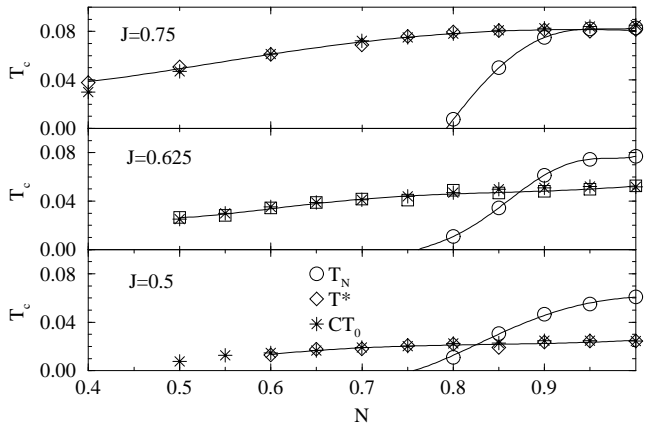}
\caption{Phase diagrams of the two-channel Kondo lattice for various
values
of $J$.  The solid lines are fits to the data.  For each value of $J$
shown
the antiferromagnetic transition occurs first (at a higher temperature)
at half filling $N=1$ (albeit very weakly when $J=0.75$).  $T_N$ falls
quickly
with doping $N<1$, with the transition becoming incommensurate near
$T_N\to0$.
Away from half filling, superconductivity occurs first above a
temperature
$T^*\geq T_c$ is roughly equal to $C T_0$, ($C=0.43$, $0.51$, and
$0.58$, for
$J=0.5$,$0.625$,$0.75$ respectively) consistent with the local nature
of the
transition.  $T^*$ should be regarded as a lower bound on the
superconducting
transition temperature $T_c$ (see text). }
\end{figure}

\end{document}